\documentclass[prl,preprint,showpacs,amssymb,superscriptaddress]{revtex4}
\begin{document}

\title{Low momentum nucleon-nucleon potentials \\
with half-on-shell T-matrix equivalence }

\author{Scott Bogner}
\affiliation{Department of Physics, SUNY, Stony Brook, New York 11794}
\author{T. T. S. Kuo}
\affiliation{Department of Physics, SUNY, Stony Brook, New York 11794}
\author{L. Coraggio}
\affiliation{Dipartimento di Scienze Fisiche, Universit\`a 
di Napoli Federico II, \\ and Istituto Nazionale di Fisica Nucleare, \\
Complesso Universitario di Monte  S. Angelo, Via Cintia - I-80126 Napoli, 
Italy}

\date{\today}

\thanks{Talk presented by TTSK at FB16 International Conference,
Taipei, Taiwan March 6-10, 2000}

\begin{abstract}
We study a method by which realistic nucleon-nucleon
 potentials $V_{NN}$
can be reduced, in a physically equivalent way, to an effective
 low-momentum potential $V^{low-k}$ confined
 within a cut-off momentum $ k_{cut}$.
Our effective potential is obtained using the folded-diagram method of
Kuo, Lee and Ratcliff, and  it is shown to
preserve the half-on-shell T-matrix. Both the Andreozzi-Lee-Suzuki
and the Andreozzi-Krenciglowa-Kuo iteration methods have been
employed in carrying out the reduction.
Calculations have been performed for the Bonn-A and Paris
NN potentials, using various choices for $k_{cut}$ such as
$2~{\rm fm}^{-1}$. The deuteron binding energy, low-energy
NN phase shifts, and the low-momentum half-on-shell T-matrix
 given by
$V_{NN}$ are all accurately reproduced by $V^{low-k}$.
Possible applications of $V^{low-k}$ directly to nuclear matter and nuclear
structure calculations are discussed.
\end{abstract}

\maketitle

\section{Introduction}
  Recently there has been much interest in studying nuclear physics problems,
particularly the two-nucleon problem, using effective field
theory (EFT) \cite{kaplan,epel00}.
The basic idea of the EFT approach is to shrink
the full-space theory  to a small-space one which contains only
the low-momentum modes. This is accomplished by integrating out the high
momentum modes, thus generating effective couplings which implicitly
contain the effects of the high-momentum modes. In this way, one can
derive a low-momentum nucleon-nucleon (NN) potential, which is specifically
designed for low-energy nuclear physics. The low-momentum NN
potentials so constructed have indeed been quite successful in describing
the two-nucleon system at low energy . \cite{epel00}

 In the present work, we would like to derive also a low-momentum NN
potential, although along a different direction.
There are a number of realistic nucleon-nucleon potentials $V_{NN}$,
such as the Bonn \cite{bonn} and Paris \cite{paris} potentials,
which all describe the observed
deuteron and NN scattering data very well. They have both low- and
high-momentum components. In fact because of the strong short range
repulsion contained in them,
their momentum space matrix elements $V(k,k')$ are
 still significant at large momentum. We would like
to  reduce such realistic potentials,
in a physically equivalent way, to certain effective low-momentum NN
potentials, $V^{low-k}$,  which have only slow momentum components,
 below a chosen cut-off momentum $k_{cut}$.

\section{Transformation method}

 The Schroedinger equation for the
full-space two-nucleon problem is written as
\begin{equation}
H\mid \Psi \rangle=E \mid \Psi \rangle;~H=H_0+V_{NN},
\end{equation}
where $H_0$ is the kinetic-energy operator for the two-nucleon system.
A model space P is defined as a momentum subspace with
$k \leq k_{cut}$, $k$ being the two-nucleon relative momentum. We want to
transform the above equation to a model-space one
\begin{equation}
PH_{eff}P\mid \Psi \rangle=E P\mid \Psi \rangle;~H_{eff}=H_0+V^{low-k},
\end{equation}
where the low-momentum effective interaction is denoted as $V^{low-k}$.
As far as the low-energy physics is concerned, we would like to have $H_{eff}$
to be physically equivalent to $H$. Specifically, this means the
requirement that the deuteron
binding energy, low-energy NN phase shifts and the low-momentum
half-on-shell T-matrix of H are all reproduced by $H_{eff}$.
The full-space half-on-shell T-matrix for $V_{NN}$ is defined as
\begin{equation}
 \langle p' \mid T(\omega) \mid p \rangle
=\langle p' \mid V_{NN} \mid p \rangle
 + \int _0 ^{\infty} k^2 dk \langle p' \mid V_{NN} \mid k \rangle
 \frac{1}{\omega -H_0(k)}\langle k \mid T(\omega) \mid p \rangle
;~\omega=\varepsilon _p,
\end{equation}
where $\varepsilon _p$ is the unperturbed energy for state $\mid p \rangle$.
The corresponding model-space T-matrix given by $V_{low-k}$ is
\begin{eqnarray}
 \langle p' \mid T_{eff} (\omega) \mid p \rangle
=\langle p' \mid V^{low-k}  \mid p \rangle
 +\int _0 ^{k_{cut}} k^2 dk \langle p' \mid V^{low-k} \mid k \rangle
 \frac{1}{\omega -H_0(k)}\langle k \mid T_{eff}(\omega) \mid p \rangle
;\nonumber \\ ~\omega=\varepsilon _p.
\end{eqnarray}
Note for $T_{eff}$ the intermediate states are integrated up to $k_{cut}$.
The boundary conditions associated with the free Green's function
are not written out, for simplicity. For p and p' both belonging to P
(i.e. both $\leq k_{cut}$), we require
\begin{equation}
 \langle p' \mid T(\omega=\varepsilon _p) \mid p \rangle
 =\langle p' \mid T_{eff}(\omega=\varepsilon _p) \mid p \rangle.
\end{equation}

Base on the Kuo-Lee-Ratcliff (KLR) folded-diagram method
\cite{klr71,ko90}, Bogner, Kuo and Coraggio \cite{bogner99} have recently
shown that the above requirements can be satisfied when the low-momentum
effective interaction is given by the folded-diagram series
\begin{equation}
V^{low-k} = \hat{Q} - \hat{Q'} \int \hat{Q} + \hat{Q'} \int \hat{Q} \int
\hat{Q} - \hat{Q'} \int \hat{Q} \int \hat{Q} \int \hat{Q} + ~...~~,
\end{equation}
where $\hat{Q}$, often referred to as the $\hat Q$-box,
is an irreducible vertex function in the sense that
its intermediate states must be outside the model space P.
 The integral sign appearing above represents a
generalized folding operation \cite{klr71,ko90}.
$\hat{Q'}$ is obtained from $\hat{Q}$ by removing terms of first order in the
interaction $V_{NN}$.

Let us outline their proof. A general term of the above T-matrix
can be written as
 $\langle p' \mid (V +V\frac{1}{e(p)}V +V\frac{1}{e(p)}V\frac{1}{e(p)}V+
\cdots) \mid p \rangle $ where $e(p)\equiv (\varepsilon _p- H_o)$.
 Note that the intermediate states
cover the entire space (1=P+Q where P denotes
the model space and Q its
complement). Writing it out in terms of P and Q, a typical term of T
is  $V\frac{Q}{e} V\frac{Q}{e}V\frac{P}{e}V\frac{Q}{e}V
\frac{P}{e}V$. Note it has three segments partitioned by two $\frac{P}{e}$
propagators.
Let us define a $\hat Q$-box as $\hat Q=V+V\frac{Q}{e}V+V\frac{Q}{e}
V\frac{Q}{e}V+ \cdot \cdot \cdot$, where all intermediate states belong
to Q. One readily sees that the previous term
is just a part of the three-$\hat Q$-box term,
and  in general we have $T=\hat Q+\hat Q \frac{P}{e} \hat Q
+ \hat Q \frac{P}{e} \hat Q  \frac{P}{e} \hat Q +\cdot \cdot \cdot$.
 By performing a folded-diagram factorization of each term with
more than one $\hat Q$-box, one can rewrite the T-matrix as
 $T=V^{low-k} +V^{low-k} \frac{P}{e} V^{low-k}
+ V^{low-k}  \frac{P}{e} V^{low-k}  \frac{P}{e} V^{low-k}
 +\cdot \cdot \cdot$.  The above result then follows.

\section{Results and discussion}

 The above $V^{low-k}$ may be calulated using
iteration methods. We have done so using both the Andreozzi-Lee-Suzuki
(ALS) and Andreozzi-Krenciglowa-Kuo (AKK) iteration methods\cite{andre96},
for both the Bonn-A and Paris potentials. We note that $V^{low-k}$ is energy
independent, and it contains less information than the full-space $V_{NN}$.
The ALS method converges to the lowest (in energy) d states of H, d being
the dimension of the model  space. (Here we have discretized the momentum
space, writing  H as a finite matrix.) In contrast, the AKK method
converges to the d states of H with maximum P-space overlaps. We have found
that the $V^{low-k}$ given by both methods are very close to each other,
an indication that the intruder-state problem \cite{suzuki93}
does not seem to be present
in our present calculation. The deuteron binding energy given by $V_{NN}$
is very accurately reproduced by $V^{low-k}$, for a wide  range of $k_{cut}$.
In Fig.1, we compare the phase shifts given
by $V_{NN}$ and those by $V^{low-k}$. They agree quite well. Empirical
phase shifts are determined up to $E_{lab}\approx 300$MeV, and they
are given by the fully-on-shell T-matrix. Hence we need to use $k_{cut}
\sim 2 ~fm^{-1}$, if we want $V^{low-k}$ to reproduce the phase shifts
up to this energy. In Fig. 2, we compare the half-on-shell
T-matrices (calculated with the principal-value boundary condition)
 given by $V_{NN}$
and $V^{low-k}$, they also agree quite well.
Note that plotted are the (k',k) matrix elements with
 $k^2=E_{lab}M/2\hbar ^2$, M being the nucleon mass.
There are a number of similarities between the model-space
reduction method used
here and the renormaliztion group method employed
in effective field theory, and it would be useful to elucidate the
connection between them. Since our method exactly preserves the
half-on-shell T-matrix, it may provide a convenient way to study
the flow equation which describes the change of the low-momentum
effective interaction with respect to the momentum cutoff.

\begin{figure}
\caption{Comparison of phase shifts from $V_{NN}$ and $V^{low-k}$.}
\label{fig.1}
\end{figure}

\begin{figure}
\caption{Comparison of half-on-shell T-matrix from $V_{NN}$ and
$V^{low-k}$.}
\label{fig.2}
\end{figure}
\newpage

To summarize,
we have reduced realistic NN potentials $V$ (Bonn-A
and Paris) to corresponding effective low-momentum potentials
$V^{low-k}$ for a model space of $k\leq k_{cut}$.
The deuteron binding energy, low-energy phase shifts and the
low-momentum half-on-shell T-matrix  given by
$V_{NN}$ are all reproduced by $V^{low-k}$.
Because of the strong short-range repulsion contained in $V_{NN}$, it is well
known that we can not use it directly in shell-model calulations of
nuclei and/or in Hartree-Fock calculations of nuclear matter; we need first to
convert $V_{NN}$ into a G-matrix, to take care of the short-range
correlations. The G-matrix so obtained is energy dependent. We have found that
our $V^{low-k}$ is a generally smooth potential (without strong short range
repulsion), and it is energy independent. It may be suitable to use
$V^{low-k}$ directly in the above
calculations, without the need of first calculating the usual
Brueckner G-matrix. This would be of interest and desirable.
We have done some shell model
calculations \cite{bogner99} in this direction, and obtained rather
encouraging results. Nuclear matter calculations using $V^{low-k}$ are
in progress.

\begin{acknowledgments}
We thank Prof. G.E. Brown for many stimulating
discussions.  This work was supported in part by the
U.S. DOE Grant No. DE-FG02-88ER40388,
and the Italian Ministero dell'Universit\`a e della Ricerca Scientifica e
Tecnologica (MURST). TTSK is particularly grateful to T.S.H. Lee
for several helpful discussions.
\end{acknowledgments}

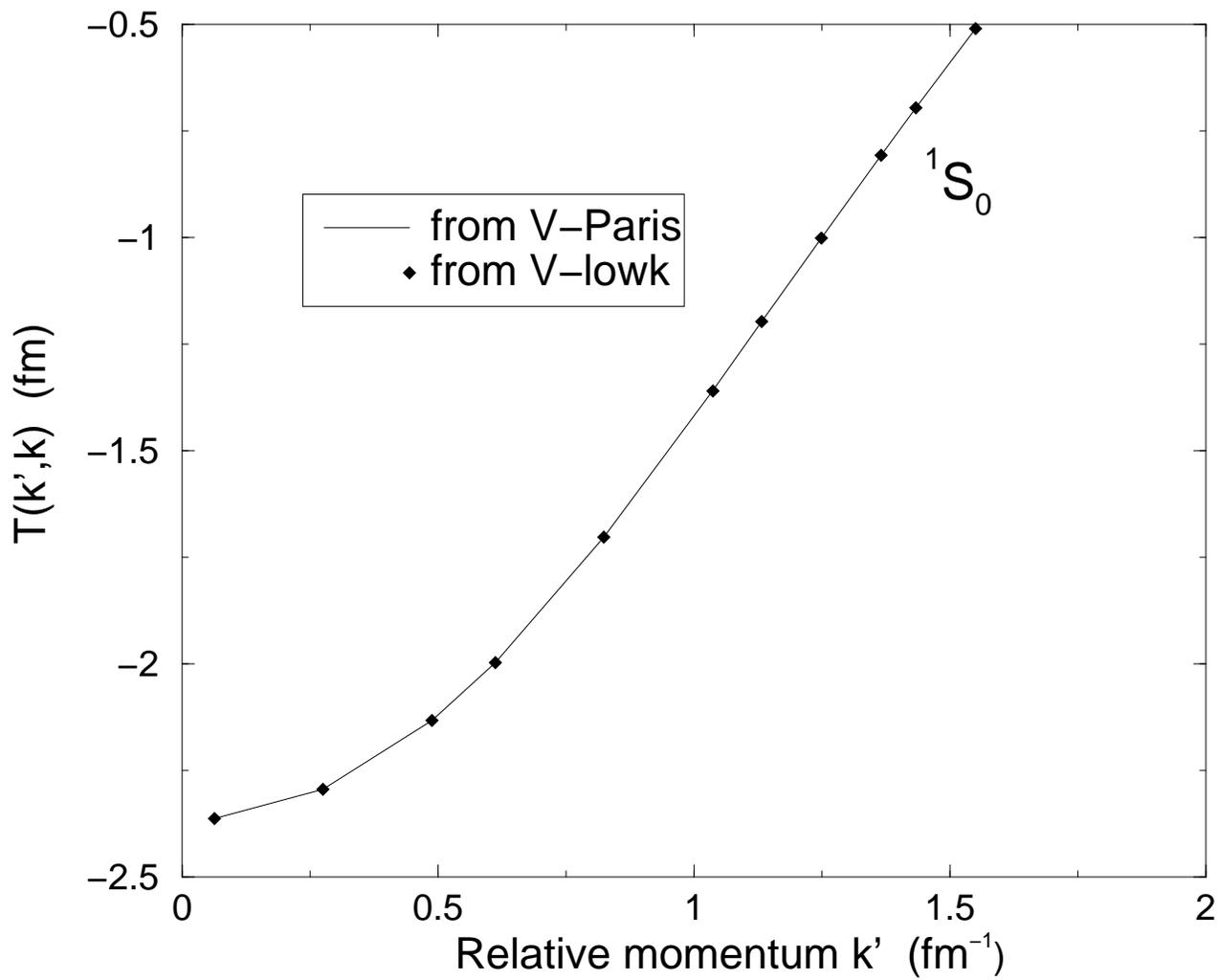
\end{document}